\documentstyle[12pt]{article}
\title{\bf Quantum cosmology of 5D non-compactified Kaluza-Klein theory}
\author{F. Darabi$^{a, b, c}$\thanks{e-mail:
fdarabi@astro.uwaterloo.ca. Send all
correspondence to
f-darabi@cc.sbu.ac.ir}, 
W. N. Sajko$^{a}$\thanks{e-mail: wnsajko@astro.uwaterloo.ca} and 
P. S. Wesson$^{a}$\thanks{e-mail: wesson@astro.uwaterloo.ca}\\
$^{a}${\small Department of Physics, University of Waterloo, Waterloo,
Ontario, N2L 3G1, Canada.}\\
$^{b}${\small Department of Physics, Shahid Beheshti University, Evin,
Tehran 19839,  Iran.}\\
$^{c}${\small Department of Physics, Tarbiyat Moallem University, Tabriz,
Iran .}
}
\begin{document}
\maketitle
\baselineskip.4in
\begin{abstract}
We study the quantum cosmology of a five dimensional non-compactified
Kaluza-Klein theory where the 4D metric depends on the fifth coordinate,
$x^4\equiv l$. This model is effectively equivalent to a 4D non-minimally
coupled dilaton field in addition to matter generated on hypersurfaces
$l=\mbox{constant}$ by the extra coordinate dependence in the
four-dimensional metric. We show that the Vilenkin
wave function of the universe is more convenient for this model as
it predicts a new-born 4D universe on the $l\simeq0$ constant
hypersurface.
\end{abstract}

\section{Introduction}

The subject of initial conditions is one of the most important questions
in cosmological models. Unlike a classical system where the dynamical
equations are solved subject to the initial conditions, for
cosmological
models there are no initial conditions external to the universe to be
considered for solving the Einstein equations. This is because there is
no time parameter external to the universe. We know that the issue of
initial 
conditions in classical cosmology corresponds to a boundary
condition problem in quantum cosmology. Therefore, it seems that the 
initial conditions should be introduced, from the outside, within some
boundary conditions. Two well-known proposals 
commonly used in the literature are the
Hartle-Hawking {\em no boundary} proposal \cite{HH1}-\cite{HH4}, and the
Vilenkin
{\em tunneling} proposal \cite{Vil5}-\cite{Vil9}. The first proposal is
that the
universe has no boundary in 4D Euclidean space and the second one states
that only the outgoing
modes of the wave function should be taken at the singular boundary of 
superspace. Some attempts to generalize these
proposals to higher-dimensional Kaluza-Klein cosmologies, to find a
reasonable explanation to the large separation of the scale of the
observed three-dimensional universe and the scale of the extra dimensions
have already been done \cite{CLOP10}, \cite{CLOP11}. In these works, the
extra dimensions were supposed to
be stable and compactified by a cyclic symmetry to
a small size. 

In this paper, we follow another approach and
investigate the quantum cosmology of a non-compactified Kaluza-Klein
theory developed by Wesson and co-workers \cite{Wes12}-\cite{Wes15}.
Unlike the usual
Kaluza-Klein theory in which a cyclic symmetry associated with the extra
dimension is assumed, the new approach removes the
cyclic condition on the extra-dimension, and
derivatives of the metric with respect to the extra-coordinate are
retained. This induces non-trivial matter 
on the hypersurfaces of $l=$
const. 

Our goal is to investigate the effect of the 
$l$-dependence of the metric on the quantum cosmology of a simple
model and obtain the relevant initial condition on the fifth coordinate.
We
find that while the Vilenkin wave function leads to probability
distribution of quantum tunneling peaked around the $l\simeq0$
hypersurface,
the Hartle-Hawking
wave function leads to large $l$ values corresponding to highest
probability for the birth of a Lorentzian universe. This leads us to
choose the Vilenkin wave function as the more convenient one for the
non-compactified Kaluza-Klein theory, since it seems unnatural that
the 4-dimensional universe was born very far from the $l\simeq 0$
constant hypersurface.

\section{Wheeler-DeWitt equation}

In this non-compactified Kaluza-Klein theory we choose the
general 5-dimensional metric as:
\begin{equation}
ds^2=-\hat{N}^2(t, l)e^{-b\sigma(t)}dt^2+\frac{\hat{a}^2(t,
l)e^{-b\sigma(t)}}{[1+\frac{1}{4}kr^2]^2}dx^i dx^i +\epsilon
e^{2b\sigma(t)} dl^2.
\label{1}
\end{equation}
Here $l$ is the fifth coordinate, and $\hat{N}(t,
l)=N(t)f(l)$, $\hat{a}(t, l)=a(t) \chi(l)$ are the $l$-dependent
separable lapse function and scale factor respectively. Also $k=0,
\pm1$ is the curvature, $b$
is a parameter, $\sigma(t)$ is a dilaton field and $\epsilon=\pm1$
which leaves the signature of the fifth dimension general. The
$l$-dependence of the 4D geometry, namely $\hat{N}(t, l)$ and $\hat{a}(t,
l)$ 
indicates that the cyclic condition on the fifth coordinate is removed.
The Ricci curvature scalar is calculated to be 
\begin{equation}
\hat{{\cal R}}=\left[-6\frac{\ddot{\hat{a}}}{\hat{
N}^2\hat{a}}+b\frac{\ddot{\sigma}}{\hat{N}^2}
-6\frac{\dot{\hat{a}}^2}{\hat{
N}^2\hat{a}^2}-\frac{3b^2}{2}\frac{\dot{\sigma}^2}{\hat{N}^2}
+3b\frac{\dot{\hat{a}}
\dot{\sigma}}{\hat{N}^2\hat{a}}-6\frac{k}{\hat{a}^2}\right]e^{b\sigma}+
6\epsilon\left[\frac{\hat{N}'
\hat{a}'}{\hat{ 
N}\hat{a}}+\frac{\hat 
{N}''}{3\hat{N}}+\frac{\hat{a}'^2}{\hat{a}^2}+\frac{\hat{a}''}{\hat{a}}\right]e^{-2b\sigma}.
\label{2} 
\end{equation}
Inserting the scalar curvature into the 5D vacuum Einstein-Hilbert action
$$
S=\int dx^5 \sqrt{-\hat{g}}\hat{\cal R},
$$
leads to the following effective action
$$
S=\int \hat{L} dt,
$$
with
\begin{equation}
\hat{L}=\hat{ 
N}\left[\frac{\hat{a}\dot{\hat{a}}^2}{2\hat{ 
N}^2}-\frac{b^2}{8}\frac{\hat{a}^3\dot{\sigma}^2}{\hat{N}^2}-
\frac{1}{2}k\hat{a}+\frac{\epsilon}{2}e^{-3b\sigma}\hat{a}^3 F(l)\right].
\label{3}
\end{equation}
Here
\begin{equation}
F(l)=\frac{f'\chi'}{f 
\chi}+\frac{f''}{3f}+\frac{\chi'^2}{\chi^2}
+\frac{\chi''}{\chi},
\label{4}
\end{equation}
and the dilaton field potential is
\begin{equation}
U(\sigma)=\epsilon e^{-3b\sigma}F(l).
\label{5}
\end{equation}
The Hamiltonian form of the action can be written
$$
S=\int (\hat{P}_a \dot{\hat{a}}+P_\sigma \dot{\sigma}-\hat{N}\hat{H})dt,
$$
where $\hat{N}$ appears as the Lagrange multiplier. The variation
of the action with respect to $\hat{N}$ leads to the Hamiltonian
constraint 
\begin{equation}
\hat{H}=\frac{\hat{P}_a^2}{2\hat{a}}-\frac{2}{b^2}\frac{P_\sigma^2}{\hat{a}^3}+\frac{1}{2}k\hat{a}
-\frac{1}{2}\hat{a}^3U(\sigma)=0.
\label{6}
\end{equation}
Since no lapse function $N(t)$ appears in the Hamiltonian, 
it is an indication of the invariance of the Hamiltonian under time
reparametrization. This means $N(t)$ has no physical
importance and we may choose the common gauge in cosmology, namely
$N(t)=1$. The Wheeler-DeWitt equation in the minisuperspace
of coordinates $0<R<\infty, -\infty<\sigma<\infty$ can then be written as
\begin{equation}
\left[-\frac{1}{\hat{a}^p}\frac{\partial}{\partial
\hat{a}}\hat{a}^p\frac{\partial}{\partial
\hat{a}}+\frac{4}{b^2\hat{a}^3}\frac{\partial^2}{\partial
\sigma^2}+k\hat{a}-\hat{a}^3U(\sigma)\right]\Psi(\hat{a}, \sigma)=0.
\label{7}
\end{equation}
Here $p$ covers the ambiguity in factor-ordering, but since we will
restrict ourselves to the semiclassical approximation omitting the first
derivatives, this factor is not
important \cite{Hal}. Then, the
Wheeler-DeWitt equation can be rewritten as 
\begin{equation}
\left[-\frac{\partial^2}{\partial \hat{a}^2}
+\frac{4}{b^2\hat{a}^2}\frac{\partial^2}{\partial
\sigma^2}+W(\hat{a}, \sigma)\right]\Psi(\hat{a}, \sigma)=0,
\label{8}
\end{equation}
where
\begin{equation}
W(\hat{a}, \sigma)=\hat{a}^2[k-\hat{a}^2U(\sigma)]
\label{9}
\end{equation}
is the superpotential. 
In the investigation of the Wheeler-DeWitt equation we need to know the
properties of the superpotential $W(\hat{a}, \sigma)$. For fixed $\sigma$
the
superpotential consists of two terms, a curvature term $k\hat{a}$ and the
term
$\hat{a}^3U(\sigma)$, where $U(\sigma)$ acts effectively as a cosmological
term. By
appropriate choices for $k$ and $\epsilon$ the superpotential may have
a maximum necessary for quantum tunneling. However, along the
line of fixed $\hat{a}$, the potential
$U(\sigma)$ has no a maximum. This would suggest that we may concentrate
on
the $\hat{a}$ coordinate as a viable dynamical variable in the
investigation
of 
quantum tunneling in the $\hat{a}$ direction, and consider the
coordinate $\sigma$ as a parameter. We will discuss this subject
in the next section. 

The relevant classical cosmology subject to quantization is a closed
universe with $k=+1$. This is because the universes with $k=-1, 0$ lead to
an infinite volume in the integration of the action and so the nucleation
probability of the universe in the quantum creation procces would be
zero.
Thus, we take a closed universe, $k=+1$. The superpotential
(\ref{9}) exhibits a barrier if the dilaton potential (\ref{5}) is greater
than zero
$$
U(\sigma)>0,
$$
or (if $\epsilon=1$)
$$
F(l)>0.
$$  
In the semiclassical approximation we want to find the most probable
initial conditions for the classical motion of the universe. This motion
is controlled by the superpotential (\ref{9}). To this end, we write 
\begin{equation}
\hat{a}=\hat{a}_0(\sigma)=\frac{1}{\sqrt{U(\sigma)}},
\label{10}
\end{equation}
and so define a surface of constant superpotential $W=0$ in the
minisuperspace. In fact, equation (\ref{9}) describes a superpotential
barrier in the $\hat{a}$ direction and equation (\ref{10}) separates the
under-barrier region $0<\hat{a}<\hat{a}_0$ from the outer region
$\hat{a}>\hat{a}_0$. From
(\ref{10}) we can also obtain an equation for $f(l)$ and $\chi(l)$:
\begin{equation}
\chi \chi'
\frac{f'}{f}+\frac{\chi^2}{3}\frac{f''}{f}+{\chi'}^2+\chi\chi''=\mbox{const}.
\label{*}
\end{equation}
The
presence of a barrier region indicates that we can consider the nucleation
of the universe through a quantum tunneling effect as discussed above.
Therefore, we proceed to consider the well-known tunneling 
condition of Vilenkin
\cite{Vil9}. Our aim is to find the approximate
analytic
solutions of the Wheeler-DeWitt equation, under the barrier and beyond the
barrier.  

\section{Wave function}

First, we show a behaviour of the ``{\em nothing state}'' for
$$
\hat{a}^2\ll \hat{a}_0^2.
$$
According to Halliwell \cite{Hal}, in order to have a regular solution
for the wave function in the limit $\hat{a}\rightarrow 0$, the wave
function
should be $\sigma$-independent because the coefficient of
$\partial^2_\sigma$ in the Wheeler-DeWitt equation (\ref{8}) diverges.
The
Wheeler-DeWitt equation then reduces to 
\begin{equation}
\left[-\frac{d}{d\hat{a}^2}+\hat{a}^2\right]\Psi(\hat{a})=0.
\label{11}
\end{equation}
Introducing the auxiliary variable
$\Gamma(\hat{a})=\Psi(\hat{a})/\hat{a}^{1/2}$ and the
transformation $\nu=\hat{a}^2/2$, equation (\ref{11}) reduces to the
modified
Bessel equation
\begin{equation}
\nu^2 d_{\nu}^2 \Gamma +\nu d_{\nu}
\Gamma-\left(\nu^2 +\frac{1}{16}\right)\Gamma=0,
\label{12}
\end{equation}
whose independent solutions are the well-known modified Bessel functions
of order $1/4$, $I_{1/4}(\nu)$, and $K_{1/4}(\nu)$. Transforming to the
old variables, we find the growing solution
$\hat{a}^{1/2}I_{1/4}(\hat{a}^2/2)$ and
the decreasing solution $\hat{a}^{1/2}K_{1/4}(\hat{a}^2/2)$ in the
$\hat{a}$ direction. To
select one of them we will impose a matching condition with the solution
(\ref{17}) for $\hat{a}\ll \hat{a}_0$. This gives the decreasing solution
\begin{equation}
\Psi(\hat{a})=\hat{a}^{1/2}K_{1/4}\left(\frac{\hat{a}^2}{2}\right),
\label{13}
\end{equation}
which is the well-known solution of {\em nothing} due to Vilenkin
\cite{Vil8}, 
and goes like $e^{-\frac{\hat{a}^2}{2}}$ for
$\hat{a}\rightarrow0$. It was obtained by Vilenkin in the limit of small
$\hat{a}$ in
the 4-dimensional model with topology $R \times S^3$ and inflation,
without a dilaton field. Usually, in 4-dimensional cosmology {\em nothing}
is the nonsingular boundary of the superspace that includes 
three-geometries given through a slicing of a regular four-geometry
\cite{Vil8}, \cite{Vil9}. In higher-dimensional cosmologies, the extra
dimensions
usually play the role of a scalar field $\sigma$ in the
equivalent four-dimensional model, such that the non-singular boundary of
the minisuperspace  is the configuration $\hat{a}=0, |\sigma|<\infty$.
This
configuration is called {\em external nothing} since the extra dimension
is assumed to be nonzero \cite{CLOP11}.

Now, consider the common Wheeler-DeWitt equation in 4D
\begin{equation}
\left[-\frac{\partial^2}{\partial a^2}
+\frac{1}{a^2}\frac{\partial^2}{\partial
\sigma^2}+a^2-a^4 U(\sigma)\right]\Psi(a, \sigma)=0.
\label{14}
\end{equation}
The WKB solution of Eq.(\ref{14}) with Vilenkin boundary condition is well
known \cite{Vil9} in the region of the minisuperspace where the potential
$U(\sigma)$ is {\em sufficiently flat}, i.e. where
\begin{equation}
\left|\frac{U'}{U}\right| \ll max\{U(\sigma), a^{-2}\}.
\label{15}
\end{equation}
In fact, by assuming the condition (\ref{15}) the wave function becomes a
slowly-varying function of $\sigma$. Therefore, one can neglect
the derivative
with respect to the dilaton field $\sigma$. Thus, $\sigma$ plays the role
of a parameter in the
Wheeler-DeWitt equation (\ref{14}), and the problem is reduced to the
one-dimensional minisuperspace model. 

In the present model, however, the
potential $U(\sigma)=\epsilon F(l) e^{-3b\sigma}$ has a strongly
asymmetric form for $\sigma<0$ and $\sigma>0$, and so the condition
(\ref{15}) does not hold in the region of the minisuperspace
where $\hat{a}^2U(\sigma)>1$ and $\sigma\ll 0$ with $b>0$.
Nevertheless, we may use an approximation to cast the
Wheeler-DeWitt
equation (\ref{8}) into a solvable equation. The main point in the
non-compactified five-dimensional model is to find the most probable
$l=\mbox{const}$ hypersurface for the 4D universe to tunnel from {\em
nothing}. On the other hand, we are interested in the tunneling in the
direction of $\hat{a}$ which is the only $l$-dependent dynamical
variable in the model. Therefore, any result about the most probable
$l=\mbox{const}$ hypersurface will be obtained due to the $l$-dependence
of
$\hat{a}$, and is expected not to be affected by the $\sigma$ dependence
of $\Psi$ since $\sigma$ is independent of $l$. So, we may solve the
Wheeler-DeWitt equation in that region of the minisuperspace where the
condition (\ref{15}) holds and deduce the result about the most
probable $l=$ const hypersurface for tunneling. This result is anticipated
not to be changed
if we solve the Wheeler-DeWitt equation in the whole minisuperspace.
In other words, to the extent that we
are concerned about finding the most
probable $l=\mbox{const}$ hypersurface ( and the dependence of $\Psi$ on
the 
$\sigma$ field is not important to us ) we may suppose the wave
function $\Psi$ to be independent of $\sigma$. This assumption is valid at
least in the positive sector $\sigma \gg 0$ with $b>0$ where the potential
$U(\sigma)$ is sufficiently flat. Therefore, we may drop the $\sigma$
derivative and the Wheeler-DeWitt equation (\ref{8})
takes the form
\begin{equation}
\left[-\frac{d}{d\hat{a}^2}+\hat{a}^2[1-U(\sigma)\hat{a}^2]\right]\Psi(\hat{a}
)=0,
\label{16}
\end{equation}
where $\sigma$ is just a parameter and the problem is essentially
identical to the one-dimensional minisuperspace model with the dynamical
variable $\hat{a}$. In this approximation, the superpotential
barrier becomes wide for $\sigma\gg 0$ which, at first glance, makes it 
inconvenient for tunneling. However, due to the function $F(l)$ we
will show that the barrier becomes sufficiently narrow for
quantum tunneling, even in the $\sigma\gg 0$ sector of the
minisuperspace. The solutions of equation
(\ref{16}) are the well-known Vilenkin tunneling wave functions
\cite{Vil9}:
\begin{equation}
\Psi_T=\exp\left(- \frac{1 -[1-U(\sigma)\hat{a}^2
]^{\frac{3}{2}}}{ 3U(\sigma)} \right), 
\quad \hat{a}^2U(\sigma)<1,
\label{17}
\end{equation}
and
\begin{equation}
\Psi_T=\exp\left(-
\frac{1}{3U(\sigma)}\right)\exp\left(\frac{-i
[U(\sigma)\hat{a}^2 -1
]^{\frac{3}{2}} }{ 3U(\sigma)} \right), 
\quad \hat{a}^2U(\sigma)>1,
\label{18}
\end{equation}
where $\Psi$ for the region $\hat{a}^2U(\sigma)<1$ (underbarrier) has the
regular behaviour $\Psi \sim e^{-\frac{\hat{a}^2}{2}}$ for
$\hat{a}\rightarrow0$
matching with the {\em nothing} solution (\ref{13}). We notice that the 
potential term
$$
U(\sigma)=\epsilon F(l) e^{-3b\sigma}
$$
plays the role of an effective 4D cosmological term whose properties
merit attention.
First, the presence of $\epsilon=\pm 1$ corresponds to positive or
negative
cosmological term for a given $F(l)$. But as was discussed
before, only the $\epsilon=+1$ case is relevant to quantum
cosmology with $k=+1$, since for $\epsilon=-1$ there is no maximum
(barrier) for the 
superpotential. Second,
the presence of the $l$-dependent term $F(l)$ indicates the
contributions of the fifth dimension to the effective 4D cosmological
term.
Third, for $b \gg 1$, if the parameter $\sigma$ undergoes 
time evolution
from negative to positive values, then a large cosmological
term would
become a small one very quickly. This is important, 
as it is in agreement with quantum tunneling and inflationary ideas where
the
very early universe might have experienced an extremely
inflationary era for $\sigma<0$ which was abruptly switched off
for $\sigma>0$. This would also be an alternative solution to the
well-known cosmological constant problem in that an initially large
cosmological term becomes small after the inflationary period. 

Considering the function $F(l)$, and (\ref{4}) and (\ref{*}), we
find that a non-zero $F$ may
be achieved by taking linear behaviours for the functions $\chi$
or $f$ with respect to $l$
\begin{equation}
f(l)=\chi(l)=\frac{l}{L},
\label{z}
\end{equation}
where a constant $L$ is introduced to preserve physical dimensions. 
The choice (\ref{z}) 
corresponds to the so-called {\em canonical metric} \cite{Wes16}, and
gives
\begin{equation}
F(l)=\frac{2}{l^2}.
\label{y}
\end{equation}
The cosmological term is then obtained as 
\begin{equation}
\Lambda\equiv U(\sigma)= \frac{2}{l^2}e^{-3b\sigma},
\label{19}
\end{equation}
with the right dimension of ({\em length})$^{-2}$. 

One may obtain the probability distribution for the Vilenkin wave
function as in \cite{Vil9}:
\begin{equation}
\rho_{\small T}\sim
\exp\left[-\frac{2}{3U(\sigma)}\right]\sim
\exp\left[-\frac{l^2 e^{3b\sigma}}{3}\right].
\label{20}
\end{equation}
This is maximized for $\sigma\gg 0$ when $l\rightarrow 0.$
This condition shows that the 4+1 dimensional universe could
have
tunneled with a large probability if the fifth dimension was
very
small, and that a small 4D universe $\hat{a}_0\ll 1$ was born on a 4D
hypersurface near to the $l\simeq0$ hypersurface\footnote{We recall that
although we have taken large values of $\sigma$ in this
approximation to the Wheeler-DeWitt equation (\ref{8}), the
condition $l\rightarrow 0$ makes it possible to narrow down the
superpotential
barrier ($U(\sigma)\gg 1$) and tunnel from {\em nothing} to a small size
universe
$\hat{a}_0=\frac{1}{\sqrt{U(\sigma)}}$. }. 

Let us now consider the Hartle-Hawking wave function for this model. In 
the presence of matter fields $\phi$ the Hartle-Hawking wave function
of the universe is obtained through the functional integral over
Euclidean 4-metrics $g_{\alpha \beta}$ ($\alpha, \beta=0, 1, 2, 3$):
\begin{equation}
\Psi(\tilde{h}_{ij}, \tilde{\phi})=\int_{\cal C} dg_{\alpha \beta} d\phi
\exp[-I(g_{\alpha \beta}, \phi)].
\label{a}
\end{equation}
Here the domain ${\cal C}$ is defined by the `` no boundary'' proposal as
all regular compact Euclidean 4-geometries ( the boundary of which is
$S^3$
with the induced 3-metric $\tilde{h}_{i, j}$ ($i, j=1, 2, 3$) ) and the
regular matter field configurations ( the value of which is
$\tilde{\phi}$ 
on the 3-manifold). However, in our model there are no extra matter
fields, 
other than $\sigma$ which appears as the dilaton field in the
5-metric. In this way, the corresponding Hartle-Hawking wave function is
given as
\begin{equation}
\Psi(\tilde{\hat{h}}_{\alpha \beta})=\int_{\cal C} d\hat{g}_{AB}
\exp[-I(\hat{g}_{AB})].
\label{b}
\end{equation}
Here $\hat{g}_{AB} (A, B=0, 1, 2, 3, 4)$ is the 5-metric, and the domain
${\cal C}$ is the class of all regular compact Euclidean 5-geometries
whose boundary is $S^3 \times R$ ($R$ denotes the fifth non-compact
coordinate), with the induced 4-metric $\tilde{\hat{h}}_{\alpha \beta}
(\alpha,
\beta= 1, 2, 3, 4)$. On the other hand, the present 5D model is
effectively equivalent to a 4D non-minimally coupled dilaton field on
$l=$const hypersurfaces. Therefore, using the 5D metric (\ref{1}), the
above integral may be rewritten in its familiar 4D form in the gauge 
$\dot{\hat{N}}=0$ \cite{HH1} as:
\begin{equation}
\Psi_H(\tilde{\hat{a}}, \tilde{\sigma})|_{l=\mbox{const}}=\int
d\hat{N}\int{\cal
D}\hat{a} {\cal D}\sigma
\exp(-I[\hat{a}(\tau), \sigma(\tau), \hat{N}]).
\label{22}
\end{equation}
Here $I$ is the Euclidean action for the model, $\tau$ is the
Euclidean time and $\tilde{\hat{a}}$ and $\tilde{\sigma}$ are the final
values of $\hat{a}$ and $\sigma$ on the 3-geometry $\tilde{\hat{h}}_{i,
j}$ and
$l=$const hypersurface. However, in practice we are usually
interested in the semi-classical approximation to the above path integral
\begin{equation}
\Psi_H(\tilde{\hat{a}}, \tilde{\sigma})|_{l=\mbox{const}} \approx
\exp(-I_{cl}(\tilde{\hat{a}}, \tilde{\sigma})),
\label{23}
\end{equation}
where $I_{cl}(\tilde{\hat{a}}, \tilde{\sigma})$ is the action for the {\em
instanton} solutions to the Euclidean
field equations. The Hartle-Hawking wave function for the model defined by
the Lagrangian (\ref{3}) is well-known and may be obtained by the
following
transformation \cite{Vil9}:
$$
\Psi_H=\Psi_T(U\rightarrow e^{-i\pi}U, \hat{a}\rightarrow e^{i\pi/2}
\hat{a}).
$$
This yields
\begin{equation}
\Psi_H|_{l=\mbox{const}}\approx \exp\left( \frac{1 -[1-U(\sigma)\hat{a}^2
]^{\frac{3}{2}}}{ 3U(\sigma)} \right),
\quad \hat{a}^2U(\sigma)<1,
\label{24}
\end{equation}
\begin{equation}
\Psi_H|_{l=\mbox{const}}\approx \exp\left(
\frac{1}{3U(\sigma)}\right)\cos\left(\frac{
[U(\sigma)\hat{a}^2 -1
]^{\frac{3}{2}} }{ 3U(\sigma)} -\frac{\pi}{4}\right),
\quad \hat{a}^2U(\sigma)>1.
\label{25}
\end{equation}
Now, the probability distribution for the Hartle-Hawking wave function as
given in \cite{Vil9} is
\begin{equation}
\rho_{\small H}\sim
\exp\left[\frac{2}{3U(\sigma)}\right]\sim
\exp\left[\frac{l^2 e^{3b\sigma}}{3}\right].
\label{20}
\end{equation}
Contrary to Vilenkin's case, this probability distribution is maximized
for $l\gg 0$ .

This condition, on the other hand, indicates that a Lorentzian universe
was born 
from a {\em mother} Euclidean universe with a large probability when 
the 4D Lorentzian hypersurface was very far from the $l\simeq0$
hypersurface.

\section{Discussion}

Usually, in quantum cosmology there is a debate on the
choice between
Hartle-Hawking and Vilenkin wave functions in concern to the issue of
inflation. It is commonly believed that the
Vilenkin wave function leads to inflation. However, for some
particular models the Hartle-Hawking wave function claims to predict a
period of inflation \cite{HH2}-\cite{HH4}. So, as far as 
inflation is concerned, there is no clear way to decide between the two
proposals.

In this paper, we have introduced a new way to compare the two
proposals in a higher-dimensional Kaluza-Klein model by using the
extra dimension. We have found that if there is a
non-compactified extra dimension, the Vilenkin proposal seems more
reasonable than the Hartle-Hawking proposal. In the Hartle-Hawking
proposal, a large value of $l$
gives rise to a rather large initial radius of the
universe, which seems unnatural.
That is, in the Hartle-Hawking proposal there is no good
justification for {\em a big 4D universe which 
was born on a constant hypersurface $l\gg0$}. However, in the Vilenkin
proposal the universe starts naturally from a small radius $\hat{a}_0$
with
a large cosmological term $\Lambda\equiv l^{-2}e^{-3b\sigma}$ ($0\ll
\sigma<\infty$) on the
$l\simeq0$ hypersurface. Although we have solved the problem in the
approximation of large positive values of $\sigma$, the result $l
\simeq 0$ hypersurface is expected to remain unchanged for the whole
domain $-\infty<\sigma<\infty$ since $\sigma$ is independent of $l$.
Therefore, in the full theory
the time evolution $\sigma(t)<0 \rightarrow
\sigma(t)>0$ will switch off the large cosmological term. This
may be considered as a solution to the cosmological constant problem.

\end{document}